# $s$-Processing from MHD-induced mixing and isotopic abundances in presolar SiC grains

S. Palmerini [*], O. Trippella, M. Busso, D. Vescovi, M. Petrelli, A. Zucchini, F. Frondini

*Dipartimento di Fisica e Geologia, Universitá degli Studi di Perugia, Italy*
*INFN, Sezione di Perugia, Italy*

**Abstract**

In the past years the observational evidence that $s$-process elements from Sr to Pb are produced by stars ascending the so-called Asymptotic Giant Branch (or "AGB") could not be explained by self-consistent models, forcing researchers to extensive parameterizations. The crucial point is to understand how protons can be injected from the envelope into the He-rich layers, yielding the formation of $^{13}$C and then the activation of the $^{13}$C$(\alpha, n)^{16}$O reaction. Only recently, attempts to solve this problem started to consider quantitatively physically-based mixing mechanisms. Among them, MHD processes in the plasma were suggested to yield mass transport through magnetic buoyancy. In this framework, we compare results of nucleosynthesis models for Low Mass AGB Stars ($M \lesssim 3M_\odot$), developed from the MHD scenario, with the record of isotopic abundance ratios of $s$-elements in presolar SiC grains, which were shown to offer precise constraints on the $^{13}$C reservoir. We find that $n$-captures driven by magnetically-induced mixing can indeed account for the SiC data quite well and that this is due to the fact that our $^{13}$C distribution fulfils the above constraints rather accurately. We suggest that similar tests should be now performed using different physical models for mixing. Such comparisons would indeed improve decisively our understanding of the formation of the neutron source.

## 1. INTRODUCTION

In their Asymptotic Giant Branch, or "AGB", phases, low- and intermediate-mass stars (i.e. those that are less massive than about $8M_\odot$) are characterized by a central degenerate core of carbon and oxygen and by two overlying shells (rich in He and H, respectively) where nuclear burning occurs alternately. Above this internal structure, an extended convective envelope separates the burning regions from a cold, dusty circumstellar environment.

Stars in these stages undergo recurrent mixing phenomena, collectively called "The Third Dredge-up", or "TDU", in which the envelope is polluted by materials normally locked below the H-burning shell, where He-burning and slow $n$-captures occur. The TDU episodes develop after the temporary activation of the He-burning shell in explosive conditions, due to sudden enhancements in the temperature named "Thermal Pulses", or "TP" s. This phase is therefore often called "TP-AGB" stage.

Thanks to TDU, freshly synthesized nuclei can be observed at the surface of TP-AGB stars. Although the observations are difficult, due to the complexity and variability of the extended atmospheres, they offer a unique tool to probe the effects of nucleosynthesis while it is still going on.

[*] Corresponding author at: Dipartimento di Fisica e Geologia, Universitá degli Studi di Perugia, Italy.
  *E-mail address:* sara.palmerini@pg.infn.it (S. Palmerini).

Among AGB stars, those in a mass interval roughly extending from 1.5 to 3–4 M$_\odot$ (depending on the opacities, hence also on the metallicity) can pollute the envelope with so much material from internal zones that the surface abundance of $^{12}$C (the primary product of He-burning) overcomes that of $^{16}$O, originally larger by typically a factor of 2. These evolved objects are then called "Carbon Stars". Their observed abundances of *n*-capture elements have been analyzed in detail, e.g., by Abia et al. (2002); well before this, the correlation between these abundances and the enrichment in carbon had been clarified by several authors, especially (but not only) from the Austin group (see e.g. Smith and Lambert, 1988)

C-stars lose mass through strong, cool winds rich in dust; hence, they can pollute the interstellar medium (ISM) with C-based grains, carrying the signature of the nucleosynthesis processes that occurred in their stellar interior. Reviews of these phenomena can be found, e.g., in Iben and Renzini (1983), Busso et al. (1999), Herwig (2005), Avila et al. (2012) and Karakas and Lattanzio (2014).

From the cool stellar envelopes, tiny solids form and are dispersed into the ISM, where they can be engulfed in contracting clouds forming new stars. The recognition that this process had occurred also for the Sun, as well as the identification of methods for separating, collecting and analyzing those precious "pieces of stars" was one of the major achievements in Ernst Zinner's career: see Zinner et al. (1988), Zinner et al. (1989) and Amari et al. (1990) as examples. In particular, a very refractory condensate that can be saved from C-rich stellar envelopes is SiC (silicon carbide), whose grains were found in carbonaceous chondrite meteorites. The large majority of SiC grains belong to the so-called *Mainstream* group, which was early recognized to be formed in C-star envelopes (see e.g. Ming et al., 1989).

Modern, sophisticated techniques of analysis now allow researchers to extract, from tiny amounts of those ancient micro-solids, information not only on the most abundant nuclei, but also on the isotopic admixtures of trace elements: see e.g. Savina et al. (2003), Savina et al. (2003) and Stephan et al. (2011). These elements, heavier than iron, can be abundantly produced by the *s*-process in AGB stars.

Such a production site is characterized by a neutron release from two main processes of α-capture. The first one, the $^{13}$C$(\alpha,n)^{16}$O reaction, was originally suggested to operate in stars by Cameron (1954). It is activated in the radiative phase between two subsequent TPs as discussed, e.g., by Gallino et al. (1998). In stellar models adopting the Schwarzschild criterion for convection, and for metallicities typical of the Galactic disk, the $^{13}$C available is completely consumed in the interpulse stage, releasing neutrons at a low neutron density ($\simeq 10^7$ cm$^{-3}$). Conversely, in the early thermal pulses of stellar models where convection is extended by some form of overshoot, there is a tendency of the available $^{13}$C to survive in part the interpulse phase, thus continuing to burn in the subsequent convective instability at a higher neutron density (Guo et al., 2012; Cristallo et al., 2015). On this point see also Lugaro et al. (2012). In addition, there is the possibility, at very low metallicities and in Super AGB stellar models, that protons be ingested directly by the intermediate convective zone from the envelope, thus giving rise to a burst of *n*-captures with a high neutron concentration, up to $10^{15}$ cm$^{-3}$ (Herwig et al., 2011; Stancliffe et al., 2011; Jones et al., 2016). As we shall see later, in this contribution we consider only the first of these possibilities, hence in our models the consumption of $^{13}$C occurs before the onset of the next thermal pulse

The second source is the $^{22}$Ne$(\alpha,n)^{25}$Mg reaction, which is at play during the convective instabilities affecting the intershell region when a TP develops. Due to its requirements for a high temperature ($T \geqslant 3 - 3.5 \cdot 10^8$ K), the $^{22}$Ne source is only marginally activated in stars below about 3M$_\odot$, called *Low Mass Stars*, or LMS, where the maximum temperatures in TPs do not exceed $3.2 \times 10^8$ K (see Straniero et al., 2003; Busso et al., 2001).

The neutron densities achieved in low-mass AGB stars range from about $10^7$ cm$^{-3}$, for the neutron release by $^{13}$C$(\alpha,n)^{16}$O, to $10^{10}$ cm$^{-3}$ for the operation of the $^{22}$Ne$(\alpha,n)^{25}$Mg - source. These neutron concentrations are typical of the so-called *s*-process, where *n*-captures are slower than most weak interactions encountered along the nucleosynthesis path. The process then consists of a series of subsequent *n*-captures and β decays, proceeding very close to the valley of β-stability.

The *s*-process is responsible for the production of about half the nuclei heavier than iron (Fe), up to bismuth ($^{209}$Bi). In particular, about thirty such nuclei (the *s*-only ones) can be produced exclusively by slow *n*-captures, being shielded against the fast decays of the *r*-process by stable isotopes. Verifying the ability of reproducing the relative solar abundances of the *s*-only isotopes represents a fundamental test on the validity of any model for slow neutron captures; this is a necessary condition, which however is not also *sufficient*. The main parameters controlling the process of *slow* neutron captures are indeed essentially limited to the neutron exposure (or the number of neutrons captured per heavy seed) and the neutron density and most existing models can be adjusted to explain the solar abundances. It is exactly for this reason that more stringent constraints, like those provided by presolar grains, proved to be of crucial help for the building of *s*-process nucleosynthesis models.

## 2. CONSTRAINTS ON THE MAIN NEUTRON SOURCE IN AGB STARS

Although the $^{13}$C$(\alpha,n)^{16}$O reaction is recognized to be the most effective neutron source for *s*-processing, the $^{13}$C abundance from CN equilibrium, left in He-rich layers by shell H-burning, is very low; it would induce negligible neutron fluxes. Hence, in order to generate neutrons in a suitable concentration, special conditions must occur. In particular: (i) a mechanism for injecting protons into the H-exhausted region must be found, so that interacting with the abundant $^{12}$C they can produce fresh $^{13}$C locally; (ii) the abundance of the injected protons in each layer must be low enough not to induce further proton captures on $^{13}$C;

indeed, this would inevitably produce large amounts of $^{14}$N, which is an efficient neutron absorber and would hamper n-captures on heavy seeds;0 (iii) although the proton concentration is small, the total amount of $^{13}$C produced must be rather large, hence the proton injection must reach down to deep layers of the He-rich zone. A $^{13}$C reservoir (or "pocket") with these characteristics was shown by Maiorca et al. (2011) and Maiorca et al. (2012) to be adequate to explain the chemical evolution of the Galaxy in s-elements, including the enhancements observed in very young open clusters of the galactic thin disk.

The condition indicated at point (iii) above requires, in particular, that the $^{13}$C-enriched layers extend over regins siugnificantly larger than $10^{-3} M_\odot$ (see Maiorca et al., 2012). Although s-processing has been an active research field for decades, the observational difficulties let these new constraints on the $^{13}$C pocket emerge only recently and not without lively discussions. These concerned, in particular, the reality of the rather small s-process enhancements observed in young stellar populations, from which the requirement that the $^{13}$C reservoir be larger than previously assumed was derived. On these topics, see e.g. the contemporary, but different, conclusions derived by Bisterzo et al. (2014) and Trippella et al. (2014). See also the observations by Drazdauskas et al. (2016) for NGC5316 (confirming enhancements for s-process elemenbts) and by D'Orazi et al. (2017) (not seeing them in other very young systems, if not for Ba).

More specifically, the excess of s-elements in young stellar populations is now systematically confirmed by the papers of the ESO-GAIA survey, made with the largest existing telescope (VLT) and with UVES (probably the best available spectrometer). It is also found by some authors using different equipment, like e.g. McWilliam et al. (2013), Mishenina et al. (2013), Marsakov et al. (2016), Tautvaišienè et al. (2016) and Drazdauskas et al. (2016). However, a unanimous agreement has not yet been reached; the effects are indeed very small, at the limits of present-day spectrscopic possibilities, and subject to possible residual uncertainties in model atmopspheres and non-LTE treatments (Reddy and Lambert, 2015) so that more data are still required for a final judgement. In a recent work by Tang et al. (2017) the scientists of the ESO-GAIA survey discuss in detail the subject of the remaining controversies on the quoted abundances of s-elements. The uncertainties are admitted and analyzed; but after that, the enhancements (also for elements different than Ba) are confirmed. In Section 5 of that paper (see also Figs. 6, 7 and 10 there) the authors explicitly state that the scenario they find agrees with a trend of s-elements in the Galaxy that anti-correlates with stellar age (as originally suggested by Maiorca et al., 2011; Maiorca et al., 2012). Given the unsurpassed quality of the instrumentation used and the accuracy guaranteed by one of the most important present-day collaboration in stellar spectroscopy, we consider as wise to accept (at least for the moment) these suggestions, necessarily requiring that the $^{13}$C pocket is more extended in mass than traditionally assumed.

Anyhow, an important, independent confirmation that the $^{13}$C reservoir had to be rather different from what had been often adopted in the past came recently from the analysis of s-elements in presolar SiC grains made by Liu et al. (2014a), Liu et al. (2014b) and Liu et al. (2015). These authors compared isotopic and elemental ratios of s-elements in individual presolar SiC grains, as measured either by themselves or by Nicolussi et al. (1997) and Barzyk et al. (2007), with s-process computations. In particular, Liu et al. (2015) performed extensive sensitivity studies by letting the main parameters of the $^{13}$C reservoir (extension, abundance by mass of $^{13}$C and $^{14}$N, etc.) vary in an exploratory way. The lack of rigid prescriptions from specific physical mechanisms allowed them to find some general features that should characterize the model results, in order to reproduce the data. In particular, they indicated that accounting for the relative trends of Ba and Sr isotopes requires $^{13}$C pockets larger than traditionally assumed (i.e. larger than a few $10^{-4} M_\odot$). Also the $^{13}$C abundance inside the reservoir must be different than commonly inferred previously, remaining rather flat and at a low concentration over most of the pocket mass. These indications become now general constraints that must be taken into account by any mixing model aiming at producing an adequate reservoir of $^{13}$C. In fact, as we shall see in the next Sections, our suggestions for the mixing of protons into the He-rich zone at TDU yield a partially-mixed layer that has the general characteristics suggested by Liu et al. (2015). Basically, the papers quoted so far, although not adopting a specific physical model, derived from their exploratory approach convincing constraints on their parameters, which must now be reproduced by any further modelling. What we aim at doing here is exactly this: discuss one specific case, among possible physically-founded models, which seems to satisfy the mentioned constraints.

Very recently Bisterzo et al. (2017) made a series of tests on the Galactic Chemical Evolution of s-elements using various extensions and forms for the $^{13}$C pocket. They delayed the analysis of young populations to a future paper, so there is not a direct comparison with the results by Maiorca et al. (2012) and Trippella et al. (2016). For the rest, they found that purely stellar and Galactic observational data can be fitted, with different combinations of the parameters, by rather different choices of the $^{13}$C reservoir. If confirmed, this indication would emphasize even more the relevance of presolar SiC data and of the Liu et al. (2014a), Liu et al. (2014b) and Liu et al. (2015) analysis in informing us on otherwise inaccessible details of the mixing required for driving s-processing.

New attempts are now underway to model the formation of the neutron source $^{13}$C. While Karakas and Lugaro (2016) still propose a parameterized approach in which a proton penetration is "assumed" without invoking a specific physical mechanism for its formation, other studies try in various ways to invoke some of the many complex dynamical processes expected to occur in AGB stars. One of them was recently presented by Battino et al. (2016); it explored the effects of gravity waves and Kelvin-Helmholtz instabilities. Previous suggestions indicating gravity waves as promising mechanisms for inducing non-convective mixing in evolved stars had come by Denissenkov and Tout (2003), Talon (2008) and Decressin et al. (2013) and others. Approaches somewhat different from those quoted above, but equally interesting, have been recently pursued by Cristallo et al. (2011) and Cristallo et al. (2015), on the basis of revisions of the physics at the border of convective zones; and similar attempts have been made by Pignatari et al. (2016). It is still premature to express quantitative judgements on such revisions, mainly because they find quite different extensions for the mixed layers where the $^{13}$C neutron source is formed and for TDU (see e.g. discussions in Pignatari et al., 2016). It is however clear that these attemps point to the same direction as ours, i.e. to the need of passing from exploratory approaches containing free parameters to more evolved types of modelling, based on some physical principles. This kind of work owes a lot to previous suggestions made, e.g., by Herwig et al. (1997) and Herwig (2000). It will be very interesting to see whether these independent paths have the capability to account for the above-mentioned constraints indicated by Liu et al. (2015). The complexities of stellar plasma phenomena make actually probable that more than one mechanism be at play in evolved stars. On these subjects see also the works by Karakas (2010) and Lugaro et al. (2012).

It is in any case clear that purely diffusive mechanisms of transport (either rotationally-induced or of thermohaline nature), which are important for other purposes (Zahn, 1992; Palacios et al., 2006; Eggleton et al., 2006; Eggleton et al., 2008; Stancliffe, 2010), should be either ineffective or completely absent below a TDU episode, where the proton penetration necessary to build the neutron source $^{13}$C must occur. In particular, thermohaline mixing cannot operate at all, because it would require an inversion of the molecular weight that is not present in He-rich layers at the required moment. There, the previous thermal pulse has actually homogenized the compostion. Moreover, according to the works by Traxler et al. (2011) and by Denissenkov and Merryfield (2011), this mixing mechanism is always very slow, even in previous evolutionary stages, where it can occur; hence it is a priori inadequate to operate in the short time intervals ($\simeq$100 yr) of TDU episodes. Concerning rotational mixing, recent research indicates that, if one excludes the rare cases of very fast rotation, it would be characterized by diffusion coefficients even smaller than those of thermohaline mixing itself (see e.g. Fig. 9 in Charbonnel and Lagarde, 2010).

In this paper we aim at showing that MHD[1]-induced transport can in principle provide a satisfactory explanation for s-element data in presolar SiC grains, as an extra property that adds to the previously found capability of accounting for s-element isotopic ratios in the solar system. In order to do this, in Section 3 we present some relevant characteristics of our model, while in Section 4 we illustrate our results, also in comparison with SiC data. In Section 5 we then discuss our findings in the light of previous attempts at explaining the SiC isotopic ratios. Finally, some preliminary conclusions are summarized in Section 6.

## 3. OUR MODELS FOR MIXING AND S-PROCESSING

Dynamo studies like those by Parker (1994) revealed that, starting from the original poloidal field of a rotating star, a toroidal field of similar strength is generated. This last is prone to various instabilities (see e.g. Spruit, 1999), and to a general buoyancy of magnetized structures that descends from the presence, in them, of an extra pressure term due to the magnetic field (Schuessler, 1977). The dynamics of buoyant magnetized domains is strongly dependent on the physics of the surrounding plasma. Normally, (e.g. in the Sun) stars are in a resistive regime, where rotation and magnetic fields are strongly coupled. In such cases the physics of the plasma is very complex and the evolution should be studied numerically, including the full MHD equations (Eggenberger et al., 2005). Below the convective envelope of an AGB star, instead, a rather uncommon situation occurs (see later Section 3.1), which allows the MHD equations to be solved exactly at least for the radial component of the motion (Nucci and Busso, 2014), yielding a simple formula for its buoyancy velocity and describing a fast (several meters per second) transport mechanism. As shown in the original paper, other effects (like the slow global rotation of the stellar structure) can be neglected as compared to magnetic buoyancy.

Here lies the most crucial part of our approach and its possibility of inducing the formation of a $^{13}$C reservoir. Let us therefore analyze this issue more carefully in Section 3.1. In Section 3.2 we shall then outline the other general assumptions at the base of our calculations (the reference stellar model, the code for s-processing, the choice of the nuclear parameters, etc.).

### 3.1. MHD-induced mixing and the formation of the neutron source

As done in Nucci and Busso (2014), we consider the buoyancy of magnetized structures (*flux tubes*), in 3D coordinates that describe the radial ($r$), azimuthal ($\varphi$) and latitudinal ($\theta$) motion. Most considerations were then shown by the quoted authors to be feasible in 2D ($r$ and $\varphi$) coordinates, due to the general toroidal geometry of the system. Assuming central symmetry, we can impose that the azimuthal component of the velocity $v_\varphi$ is not dependent on $\varphi$. Moreover, the toroidal component of the magnetic field is perpendicular to the radial velocity $v_r$ (the field is solenoidal). As we shall see soon, this specific property of the magnetic "engine" is crucial in finding simple exact solutions. In the relevant stellar layers (see Fig. 1, panel a) we find that the density follows a law $\rho \propto r^k$, where k is a negative number, with a modulus much larger than 1. In this particular situation, Nucci and Busso (2014) showed that the MHD equations (i.e. Eqs. (1)–(5) in that paper) simplify drastically. In particular, the continuity equation decouples from the rest of the treatment, and can be written in the form:

$$\frac{\partial \varrho}{\partial t} + \frac{\partial(\varrho v_r)}{\partial r} + \frac{1}{r}\frac{\partial(\varrho v_\varphi)}{\partial \varphi} + \frac{\varrho v_r}{r} = 0. \qquad (1)$$

---

[1] Magneto-hydrodynamics.

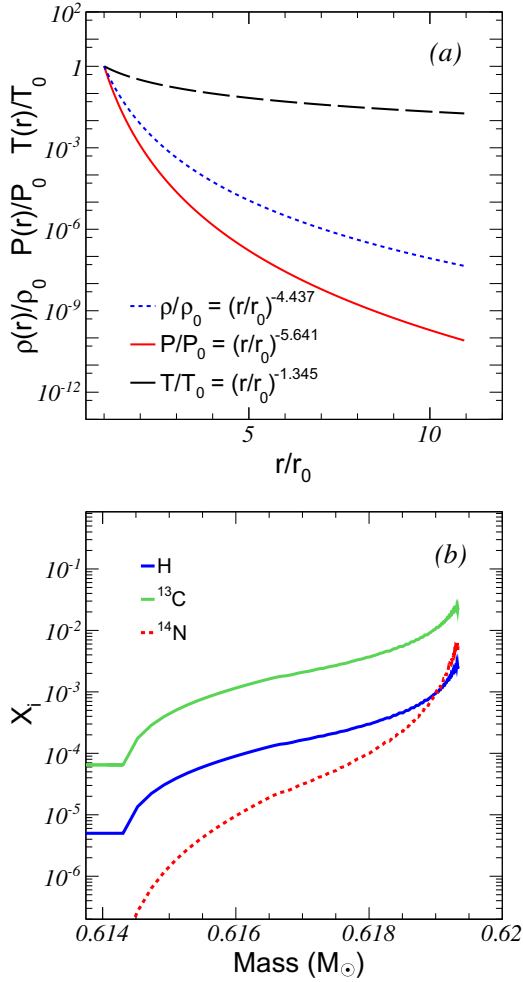

Fig. 1. Panel (a). The physical structure of the He-rich layers in an AGB star of 1.5M$_\odot$ and solar metallicity at the 6th TDU. The model was recalculated by us for this paper with the Scwarzschild criterion for convection, with the code by Straniero et al. (2003). We choose this specific structure for illustration purposes, because it represents a good average situation, suitable to desciribe the "typical" conditions on the TP-AGB. As shown, the density and the pressure decrease very rapidly as a function of the stellar radius (actually, little mass in contained in this zone, $\Delta M \simeq 0.01 M_\odot$). Panel (b). The proton profile we obtain below TDU, at the same pulse as in the first panel, together with the ensuing abundances of $^{13}$C and $^{14}$N after hydrogen reignition. The abundance of $^{13}$C dominates over that of the neutron poison $^{14}$N, limiting the latter's effects in neutron filtering; moreover, the $^{13}$C profile is more extended and flatter than provided by models based on more traditional, exponentially-decreasing proton penetration. (A color version of this figure is available in the online journal.)

Since $\frac{\partial v_\varphi}{\partial \varphi} = 0$ and the density is a known function of $r$ only ($\rho \propto r^k$), $v_r$ separates from $v_\varphi$. Eq. 1 then becomes an ordinary differential equation that can be easily solved, yielding:

$$v_r(r,t) = \Gamma(t) r^{-(k+1)}, \qquad (2)$$

This is in fact a simplification of a multi-D problem into a simple one-dimension problem. The possibility to do that is critically dependent on assuming that rotation is slow enough not to modify the circular symmetry of the star over the time scale of our interest. This simply means that rotational effects must be negligible over the duration of a TDU episode, during which we want protons to penetrate into the He-rich zone. Since such a duration is short ($\simeq 100$ years), no rotational distortion is actually expected on the basis of the common rotational velocities of low mass stars, so that this constraint is virtually always satisfied.

As mentioned, another critical condition occurring in these layers is the rapid drop of the density with radius. Actually, the region of interest is rather wide in radius (from 1 to several solar radii) but contains very little mass and this explains why dissipative terms dependent on the presence of matter and charge are in general small. Also the magnetic diffusivity is negligibly small in these conditions. Hence, in the simple geometry adopted, the induction equation (number 3 in the quoted paper) simplifies to a linear, first-order, partial differential equation in $B_\varphi$, whose general solution can be written as:

$$B_\varphi = \Phi(\xi) r^{k+1}, \qquad \xi = -(k+2)w(t) + r^{k+2} \qquad (3)$$

As is common with sets of (partial) differential equations, the *general* solution contains *mathematically arbitrary* functions. In particular, the functions $w$ and $\Phi$ in Eq. (3) and $\Gamma$ in Eq. (2) have a single link between them ($w$ must be the time integral of $\Gamma$, as shown by Nucci and Busso, 2014). Hence we have actually two arbirary and independent functions of time in our solution: let's say they are $\Gamma$ and $\Phi$. Their arbitrariety is an indication that the solution is extremely general: it will continue to exist for whatever choice of the two functions we may want to adopt.

This makes the solution itself very robust, free from model-dependencies other than those contained in the initial hypotheses already discussed. One can then notice that the basic stellar quantities on which our treatment depends (density, temperature, pressure, rotational velocity and magnetic field) vary on time scales that are extremely long as compared to the duration of an individual TDU phenomen, during which our process of mixing must be active. We are therefore interested to a behavior in which the mixing velocity is time-independent. In it, the mentioned functions will assume some specific values $\Gamma(t_0)$, $\Phi(t_0)$, simply acting as coefficients of the functional dependence on $r$ of the buoyancy velocity and of the magnetic field. In our *physical* application of the *mathematical* solution, the arbitrariety of the functions must also be handled so that the involved quantities are expressed in meaningful units and have values in accordance with observations. This implies that $\Gamma(t)$ must be chosen to have dimensions [cm$^{(k+2)}$ s$^{-1}$], while $\Phi$ must be measured in units [Gauss cm$^{-(k+1)}$] (as a consequence, $w$ must have dimensions [cm$^{k+2}$], same as for $\xi$). For the details of the calculations see again Nucci and Busso (2014).

In the family of solutions fulfilling all the requirements, for our purposes we can limit ourselves to those describing a fast transport, as done originally by the quoted authors; this is dictated by the mentioned need of having processes effective enough to occur in the short time interval available at a TDU. In particular, if $r_0$ is the radius of the layer at which MHD becomes quasi-ideal (so that the buoyancy can start), one can adopt:

$$w(t) = \omega r_0^{(k+2)} t \tag{4}$$

and

$$\Phi(\xi) = A \cos[r_0^{-(k+2)}(w_0 + r^{k+2})], \tag{5}$$

where $w_0$ is the value of the function $w(t)$ at the time of the chosen TDU episode, $t = t_0$. This yields a purely-expansive solution for $v_r$, i.e. for the buoyancy of magnetized structures:

$$v_r(r) = v_{r,0}\left(\frac{r}{r_0}\right)^{-(k+1)}, \tag{6}$$

$$B_\varphi(r, t_0) = B_{\varphi,0} \cos(\omega t_0 + r_0/r)\left(\frac{r_0}{r}\right)^{-(k+1)}. \tag{7}$$

Here $v_{r,0}$ is the value of $v_r$ at $r = r_0$. As discussed in Nucci and Busso (2014), $v_{r,0}$ and $B_{\varphi,0}$ can be approximately estimated from the values of the velocity and of the field at the base of the convective envelope, $v_{r,e}$ and $B_{\varphi,e}$. Observational constraints suggest that the magnetic field there should be of the order of $2 - 3 \times 10^4$ G (see also Palmerini et al., 2017). Concerning $v_{r,0}$, similar arguments lead to values between 10 m/s below TDU (Trippella et al., 2016) and 100 m/s at the envelope border, when the H-shell is active (Palmerini et al., 2017). These constraints were derived from previous works: see also discussions in Nucci and Busso (2014). Although the specific choice for $v_{r,e}$ is not very critical for the $^{13}$C pocket formation (it is sufficient that it is large enough to permit proton penetration even in the short time interval available) its value was fixed previously for other puroposes (the explanation of the solar s-process distribution and the reproduction of post-AGB abundances). In the application to the problems of interest here, we cannot make futher changes: the approach can be either correct or wrong, but it is not further adjustable. Here we have applied the model by Trippella et al. (2016) for the formation of the $^{13}$C pocket also to presolar grain data; and they seem to confirm the validity of the choice. We also underline that the *arbitrary* functions mentioned above, which one has to specify in order to get a physically meaningful solution on the AGB out of a set of *mathematical* equations, do not allow us to make exploratory choices of parameters exclusively for the $^{13}$C pocket, as instead done often in other approaches to s-processing. These last scenarios intentionally avoid a reference to a physical mechanism, in order to offer general guidelines to subsequent models: see e.g. the clear discussions made by Liu et al. (2015) and Lugaro et al. (2014). This is the fundamental difference between our procedure and those previous works. The type of computations presented here, much like those in Cristallo et al. (2015), can therefore be referred to as being first examples of *physically based* approaches. Quite obviously, this does not dimish the relevance of the previous efforts; to them we owe a basic understanding of the conditions to be met, without which our present upgrades would not have been possible.

Although here $v_r$ does not depend explicitly on $B$, $v_\varphi$ has a complex dependence on the field. Hence, while Eq. (1) may appear as purely dynamical, the mechanism is actually magneto-hydrodynamical. The same simple form obtained for Eq. (1) would indeed not hold in other cases, i.e. for fields not perpendicular to $v_r$. The process then looks very similar to the simplest formulation discussed by E.N. Parker for the solar wind (Parker, 1960). The wind itself remains obviously magnetized, albeit Parker's solution does not explicitly show this.

We remind in any case that the adopted solution represents a *fast* case, out of a wealth of MHD mechanisms that can act in red giants; while they should not be relevant for mixing at TDU, they might have other general effects on AGB physics, e.g. on stellar oscillations, which deserve better scrutiny in the future.

In the scenario we outlined above the process of mixing is described by an advection to the envelope of material from the layer $r_0$ in the radiative zone, while at lower values of $r$ a more complex, resistive physics characterizes the core. Magnetic tension then can keep the magnetized structures isolated from the environment for times long enough to allow them to reach the envelope. As mentioned above, the values of $B$ required for such a mechanism to work are of the order of $10^4$ Gauss at the base of the envelope, similar to what is found in the Sun. This implies the existence, at the surface, of fields with intensities from one to a few Gauss (Nucci and Busso, 2014). Recent measurements on SiO and OH masers do find that such moderate magnetic fields exist in single AGB stars (Vlemmings, 2014). These fields were suggested to be generated by a differential rotation between the core and the convective envelope (Blackman et al., 2001) and later shown to require only a moderate energy supply from the envelope convection to be maintained (Nordhaus et al., 2008). Note that fields like those required at the base of the envelope, or larger, were inferred in red giants from Kepler's observations (Fuller et al., 2015). Due to the above considerations, we can be rather confident that field values in the range required for maintaining our process do exist naturally in AGB stars, so that the required value of $B_{\varphi,0}$ does not actually introduce any real limitation on our results. We recall that sometimes white dwarfs show instead much stronger fields. They are not related to what we are discussing here; actually, a recent study from Briggs et al. (2015) suggests that they probably derive from previous common-envelope evolutionary phases in binary systems.

The conditions necessary to establish the transport were shown by Trippella et al. (2016) to exist in He-rich zone below the inner border of envelope convection at TDU. As shown quantitatively in that paper, the consequences are the buoyancy of He-shell materials to the envelope and the injection of envelope materials into the underlying radiative zones.

The ensuing proton profile obtained is illustrated (in a rather typical case) in Fig. 1 (panel b). See equations from 1 to 17 in Trippella et al. (2016) for a detailed description of this mixing process.

### 3.2. The reference codes for the stellar structure and neutron captures

We mentioned already that, among modern models trying to explain the formation of the $^{13}$C reservoir through physical mechanisms, an important role is played by re-evaluations of the physics at the inner convective border. They include possible downward extensions of the border itself due, e.g., to overshooting or to rotationally-induced mechanisms. Since we want to explore separately the consequences of our MHD approach, we adopted stellar models not including these phenomena. In particular, our MHD mixing scheme was applied, as a post process, on the stellar structures computed with the FRANEC code using a pure Schwarzschild criterion for describing the extension of convective zones. Models of this kind were illustrated, e.g., by Straniero et al. (2003).

In the framework described by those authors, a rather extended set of full stellar evolutionary calculations had been performed by us for LMS with metallicities from solar to one-tenth solar, and applied to several problems of AGB nucleosynthesis (see e.g. Busso et al., 1999; Busso et al., 2001; Abia et al., 2002; Wasserburg et al., 2006; Maiorca et al., 2011; Maiorca et al., 2012). If one excludes the layers immediately adjacent to the convective border and the extension of TDU, in those models the structure of the radiative layers of AGB phases where H and He burning occur in shells are very similar to those illustrated by Cristallo et al. (2011), as noticed in Palmerini et al. (2017).

Of the original models, sometimes dating back to more than 15 years ago, we recomputed selected cases (namely, those for 1.5, 2 and 3 $M_\odot$ stars, with metallicities from solar to one-third solar). Mass loss rates were chosen from the Reimers' parameterization. In the cases with the lower masses (1.5 and 2 $M_\odot$) we paid attention to the final stellar magnitudes: the free parameter $\eta$ of the mass loss formula was chosen so that the final magnitudes fulfill the constraints posed by Guandalini and Cristallo (2013) and the minimum mass for C-stars at solar metallicity be $1.5 M_\odot$ (Groenewegen et al., 1995). In particular, we adopted $\eta = 0.75$ for $1.5 M_\odot$ models, and $\eta = 1$ for the $2 M_\odot$ cases. With these choices the final magnitudes never exceed the value $-5.3$. For the $3 M_\odot$ cases we unfortunately do not have clear constraints on the luminosities. Indeed, these stars evolve on the AGB emitting mainly infrared radiation, so that the limits on their luminosities are still beyond our reach. One must therefore be conscious that mass loss represents the highest uncertainty in the present AGB scenario and also in our modelling. In the $3 M_\odot$ cases we adopted two different choices: $\eta = 1.5$ and $\eta = 2.5$: both choices will be illustrated in the figures.

In order to make clear what can be obtained by the models independently of our assumptions for mass loss rates, we choose here to show the isotopic ratios not only for the envelope material, affected by stellar winds at the assumed (very uncertain) rates, but also in the so-called G-component (Gallino et al., 1990). We remind that this is the material cumulatively mixed to the envelope, hence weighted on the TDU efficiency, but not diluted with the original envelope composition. It represents therefore the maximum modification of the isotopic abundances possible for s-process nucleosynthesis; any specific (and uncertain) choice for the mass loss rate provides a different level of dilution between this component and the original envelope. For the initial abundances, we scaled to the appropriate metallicity our reference solar-system abundances, taken from the compilation by Lodders and Palme (2009).

Concerning neutron captures, we adopt our post-process code NEWTON, which is an evolution of the one originally used in Busso et al. (2001) and later continuously updated in Wasserburg et al. (2006), Maiorca et al. (2011), Maiorca et al. (2012) and in Trippella et al. (2014), Trippella et al. (2016). The choices for the neutron capture cross sections were taken from the last version of the KADONIS database (see Dillmann et al., 2006; Dillmann, 2014).

## 4. RESULTS

Based on the approach outlined above (see details in Trippella et al., 2016; Palmerini et al., 2017), we computed the proton penetration into the He-rich layers and the consequent formation of the $^{13}$C neutron source for the mentioned models of 1.5–3$M_\odot$, in the metallicity interval from solar to 1/3 solar. Fig. 1 (panel b) presents the results of such calculations in a typical case, which yields a large $^{13}$C reservoir (several $10^{-3}$ solar masses), with a distribution quite different from the exponentially-decreasing trend commonly assumed in several previous approaches. It is for example different from what was adopted in recent analyses of the s-process, like the ones by Käppeler et al. (2011) and Bisterzo et al. (2014). In particular, high abundances of $^{14}$N are formed only in a very thin layer adjacent to the envelope, while for the rest of the large affected zone $^{13}$C dominates, albeit maintaining a rather small abundance. The extension of the polluted zone depends on the specific stellar model and evolutionary phase. It can be computed from the outputs of the evolutionary calculations by estimating where, in the $\rho, T$ distribution below the envelope, the conditions for quasi-ideal MHD hold at each TDU. The depth of this zone amounts to about $5 \times 10^{-3}$ $M_\odot$ for the first pulses and the lowest masses considered (see Fig. 1b) and then decreases for increasing initial mass and thermal pulse number. This corresponds to the fact that the same extension of the intermediate He-rich buffer is a decreasing function of the mass of the H-exhausted core, $M_H$, which actually controls many details of the AGB physics (Iben and Renzini, 1983; Straniero et al., 2003). The mentioned trend is also illustrated in Fig. 2 for two representative cases ($M = 1.5 M_\odot$, [Fe/H] $= -0.3$ and $M = 3.0 M_\odot$, [Fe/H] $= -0.5$). In the first two panels we show the time evolution of the position in mass for the He-shell, for the H-shell and for the convective envelope border (from bottom to top in the figures). In the three panels below we present zooms of the above trends, illustrating the succession of the pulse-interpulse phases for: (i) an early He-shell instability of a $1.5 M_\odot$ star ($\eta = 0.75$, [Fe/H] $= -0.3$), (ii) an early He-shell instability of a $3 M_\odot$ star ($\eta = 1.5$, [Fe/H] $= -0.5$); and (iii) a later instability of the same $3 M_\odot$ model ($\eta = 1.5$, [Fe/H] $= -0.5$). The figure should make clear that the layers bracketed by the inner border of convection at TDU (above) and by the position of the He-shell (below) shrink when $M_H$ increases. As a consequence, also the part of those layers where proton penetration occurs will shrink; in the last pulses of the most massive cases considered here the depth of the $^{13}$C pocket reduces to about $1.5 \times 10^{-3} M_\odot$.

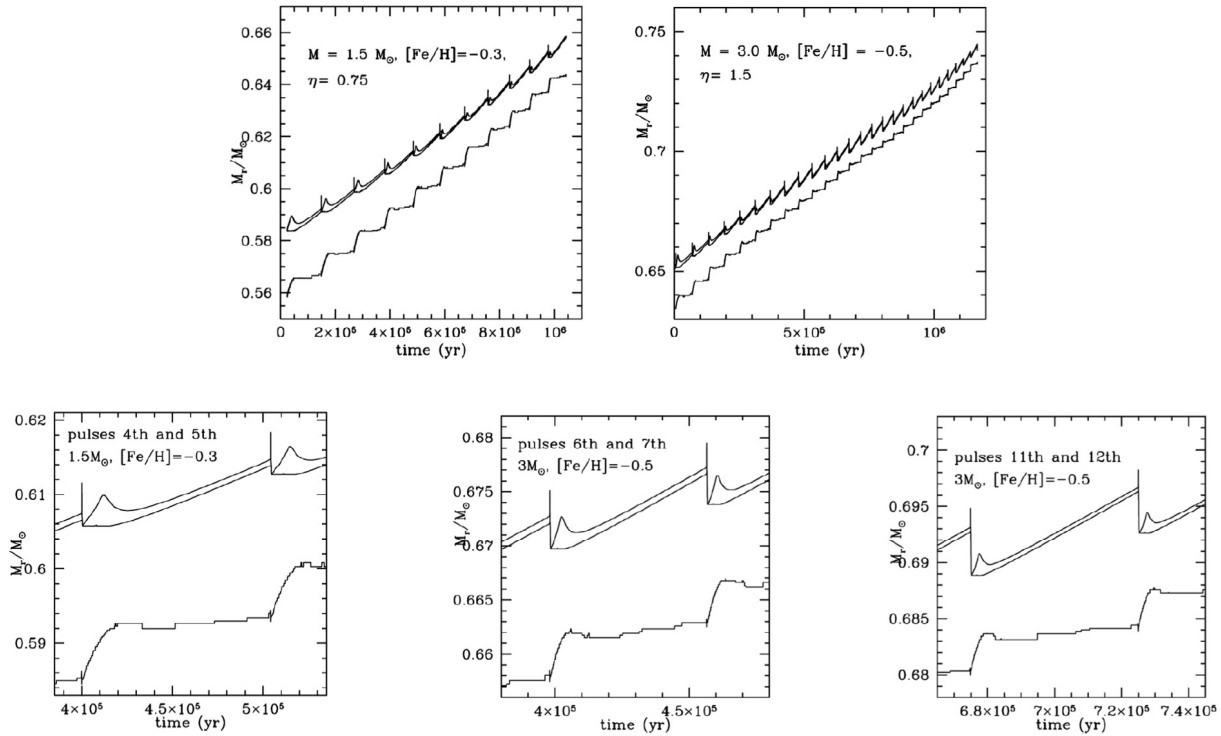

Fig. 2. Some features of the stellar models used for this work. The top panels show the evolution in time of the position of the He shell, of the H shell and of the inner border of the convective envelope, during the thermally-pulsing evolutionary phase for two selected models: that of a $1.5 M_\odot$ star of half-solar metallicity and that of a $3 M_\odot$ star of one-third-solar metallicity. The choices of mass loss rates correspnd to $\eta = 0.75$ and 1.5, respectively (see text). The bottom panels present zooms illustrating how the mass included in the He-rich zones decreases both when the number of thermal pulses increase in a given star, and for increasing initial stellar mass.

### 4.1. The general scenario emerging from our models

In the framework of the models outlined above, Trippella et al. (2016) verified that s-processing can account for the solar main s-process component within the experimental uncertainties.

The above control was made in the light of previous discussions by Maiorca et al. (2012) and Trippella et al. (2014), according to which the results of a full Galactic Chemical Evolution calculation for s-elements can be well mimicked by a suitable individual model of a specific metallicity and (low) mass, acting as a *weighted average* of the many contributions that were integrated in the solar abundance distribution. Indeed, one has to conser that, while solar abundances are a result of the ejecta of all stellar masses over the about 9 billions of years of galactic evolution that preceeded the formation of the Sun, the Salpeter's initial mass function (IMF) favors stars of low mass: and among them, those of 1.5–2$M_\odot$ share the properties of being able to undergo TDU, and to evolve fast enough ($\tau_{evol} \leqslant 2$ Gyr) to contribute repeatedly. It is also possible (Gallino et al., 1998; Trippella et al., 2014) to identify an average metallicity (depending on the average mass of $^{13}$C consumed in every pulse-interpulse cycle) giving the domi-nant contribution: the larger is the pocket in mass, the larger is this "effective" metallicity (see the quoted papers for a discussion of this point).

The result of the mentioned test is shown in Fig. 3, at the end of the evolution (i.e. after 11 TPs), for a 1.5$M_\odot$ model with a metallicity [Fe/H] = −0.15. The overabundance pattern illustrated in the figure for s-only nuclei (black dots) is very flat, showing that this model is suitable to produce n-capture nuclei in solar proportions (and quite efficiently, as the production factors in the He-shell exceed $10^3$). The other symbols show the contributions to nuclei non-exclusively produced by the s-process; their previously-expected s-process contributions are indicated in the labels.

Stars in the mass range we consider are plausible candidates to have formed C-stars with SiC grains in their envelopes, contributing to the samples now available from pristine meteorites. Out of these models, we should select those that really become C-rich. However, this selection is bound to be strongly model-dependent. A parameter-free combination of the C-enrichment by TDU episodes and of the mass loss rates affecting the producing stars is indeed still impossible. As mentioned (see Section 3), our assumptions for the mass loss rates are guided only by rough constraints on the luminosities and only for the lower masses.

With our choices, we find that the 1.5$M_\odot$ model of solar metallicity achieves the C-rich phase only for last two thermal pulses; this essentially agrees with indications by Groenewegen et al. (1995). Since TDU increases in efficiency for decreasing metallicity, in what follows models of 1.5M$_\odot$ can significantly contribute to the presolar C-star inventory only at the lowest metallicities explored (1/3 solar). In any case, all this heavily depends on the critical assumptions made for mass loss rates. It is for this reason that in the following figures we shall present the shifts in isotopic ratios induced by nucleosynthesis both for the specific choice of mass loss efficiency and for the "G-component" of the corresponding AGB model.

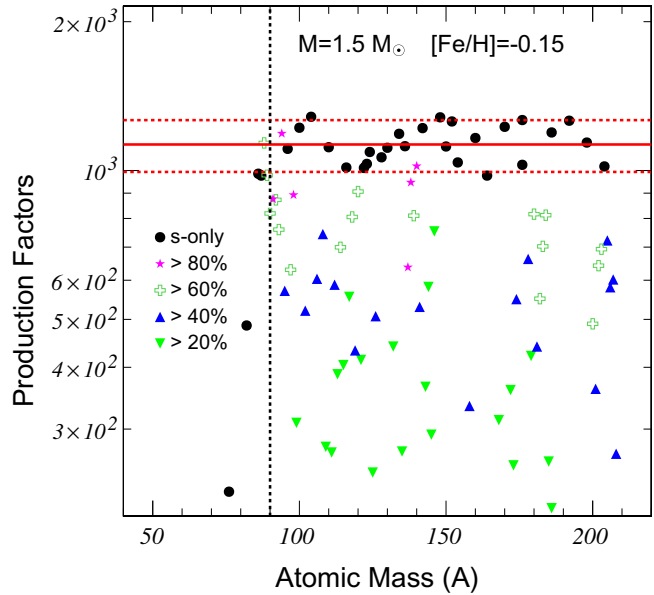

Fig. 3. The distribution of neutron capture nuclei in the He-shell for a 1.5M$_\odot$ model with $[Fe/H] = -0.15$. The ordinate shows the production factors with respect to solar abundances. Note that they have a small spread around a constant value, independently of the atomic mass number. This fact demonstrates that the solar-system distribution is reproduced rather well by the nucleosynthesis in this star. Then, as shown e.g. Maiorca et al. (2012) and Trippella et al. (2014), this star also mimics an average behavior weighted over the Initial Mass Ficntion and therefore corresponds to a sort of *Galactic average*. (A color version of this figure is available in the online journal.)

At the upper end of the mass interval considered, it is known that C-rich stars exist up to about 4M$_\odot$, as discussed, e.g., by Bergeat et al. (2002). For still larger masses, the occurrence of H-burning at the base of the envelope (Hot Bottom Burning, or HBB) inhibits the C-star formation (see e.g. Groenewegen et al., 1995). We therefore expect that mainstream SiC grains are formed only in the limited mass range from 1.5 to 3–4M$_\odot$.

A general result (Palmerini et al., 2017) is that, outside this interval, stars of very low mass ($M \leqslant 1.5M_\odot$) play a crucial role for the abundance anomalies measured in presolar corundum dust, Al$_2$O$_3$. This was already an expectation simply on the basis of the Inital Mass Function; however, it was also shown by Palmerini et al. (2011) that very low masses are favored also from the point of view of the effectiveness of deep mixing (see in particular Fig. 11 in that paper). On the other hand, it has been recently pointed out that HBB stars might contribute to the Early Solar System inventory of short-lived radioactivities, especially for $^{26}$Al (Wasserburg et al., 2016). A result that is of relevance for this specific point is the one discussed in Maiorca et al. (2012), according to which the effectiveness of s-processing must decrease with increasing stellar mass. This suggestion has some general astrophysical support if the formation of the neutron source comes from MHD processes. First of all, stellar magnetic activity seems to be a specific property of LMS and to decrease for higher mass objects: see e.g. the discussion by Schrijver and Zwann (2000). The dying out of magnetic activity should also induce a reduction in the mass affected by magnetic buoyancy and hence a reduction in the $^{13}$C-pocket size. Furthermore, in the same direction plays the fact that the intershell region is a decreasing function of the core mass $M_H$ (see Fig. 2). This means that the pocket shrinks, in a given star, for increasing pulse number: it also shrinks for a given pulse in stars of increasing mass (see e.g. Iben and Renzini, 1983; Busso et al., 1999; Straniero et al., 2003; Herwig, 2005).

On the above considerations we can guess reasonably that there will be a maximum mass (say, 4 or 4.5M$_\odot$) where the $^{13}$C pocket should gradually disappear. This crucial mass range separating LMS from IMS will be examined in some detail in a forthcoming paper.

### 4.2. Specific results for the isotopic abundances of SiC grains

With the database of stellar evolution and nucleosynthesis computations thus constructed, we tried to reproduce the relative distributions of s-process isotopes and elements, as measured in SiC grains. We used for comparisons the sample of grains analyzed by Liu et al. (2015) and the data about *Mainstream* SiC grains present in the on-line Presolar Grain Database of the Washington University in St. Louis (WUSTL, Hynes and Gyngard, 2009 and subsequent updates), which includes the measurements by Savina et al. (2003), Barzyk et al. (2007), Marhas et al. (2007) and Liu et al. (2014a), Liu et al. (2014b). This repository is yet another very useful tool for nuclear astrophysicists made available by Ernst Zinner and his colleagues.

For Zr isotopes we also analyzed the measurements by Akram et al. (2015). However, they are hardly comparable with the rest of the data. Indeed, they refer not to individual presolar grains, but to bulk meteoritic samples (Calcium Alluminum Inclusions, chondrites, etc.). Due to this fact, the anomalies are enormously diluted and the effects measured are extremely small (epsilon units, i.e. parts per ten thousands, instead of tens or hundreds of parts per mill). Due to this reason, those measured anomalies (also having large error bars) cannot provide any real constraint to our models, being essentially compatible with zero at our level of accuracy.

Since the rest of the sample contains data with very different measurement precision, we report in a darker color the grains having error bars that do not exceed a certain limit. Normally, this limit is set to 70% of the measurements themselves; in the case of Sr, with larger error bars, we chose instead a limit of 100%, to have a sufficient number of data points.

Examples of the comparisons between our predictions for isotopic and elemental ratios, pulse after pulse, and the measured data are included in Figures from 4–10. The model lines are characterized by dots that represent the various TPs: open symbols refer to the O-rich phase of the evolution, filled symbols to the C-rich phase. We remind that, although SiC grains are obviously carbon-based solids, recent work on non-equilibrium chemistry in circumstellar envelopes seems to permit, in principle, the formation of both O-rich and C-rich molecules for virtually any composition of the environment, thanks to many complex phenomena, including the photo-dissociation and re-assembly of previously formed compounds. According to Cherchneff (2006), it seems reasonable to accept that a two-component (silicate-carbon) dust may form when the C/O ratio is sufficiently high, even before it formally reaches unity, along the AGB evolution. These suggestions, however, have for the moment only limited observational confirmations. Olofsson et al. (1998) did actually find C-rich compounds like the HCN molecule in O-rich environments. Nevertheless, the formation of C-bearing refractory dust particles like SiC grains is probably more difficult and still needs to be proven through real observations of the mid-infrared SiC features in O-rich envelopes. In general, the studies of condensation sequences out of equilibrium are far from being clarified and are still highly debated. With these warnings in mind, we include in our model curves also the TPs before the envelope formally reaches C/O > 1, but (as mentioned above) marking clearly the C-rich phase by using different symbols.

We choose here to show, for example, a limited (but hopefully sufficient) set of comparisons between model predictions and measurements, plotting some of the most representative isotopic ratios for light as well as for heavy s-elements (often referred to as *ls* and *hs*, respectively).

We included strontium isotopes (Fig. 4) because a production of strontium stronger than in previous models is a characteristic of the calculations by Trippella et al. (2016). For it, we adopted as a reference nucleus $^{86}$Sr, (as done by Liu et al., 2015). We do not present $\delta$ values for isotopic ratios including $^{84}$Sr, because it is a *p*-only nucleus, which is only destroyed in *s*-processing. This fact makes its envelope abundance a direct consequence of the assumed initial one. Since the abundances of *p*-nuclei in presolar stars are uncertain, they would add a potentially unknown systematic error in the model curves.

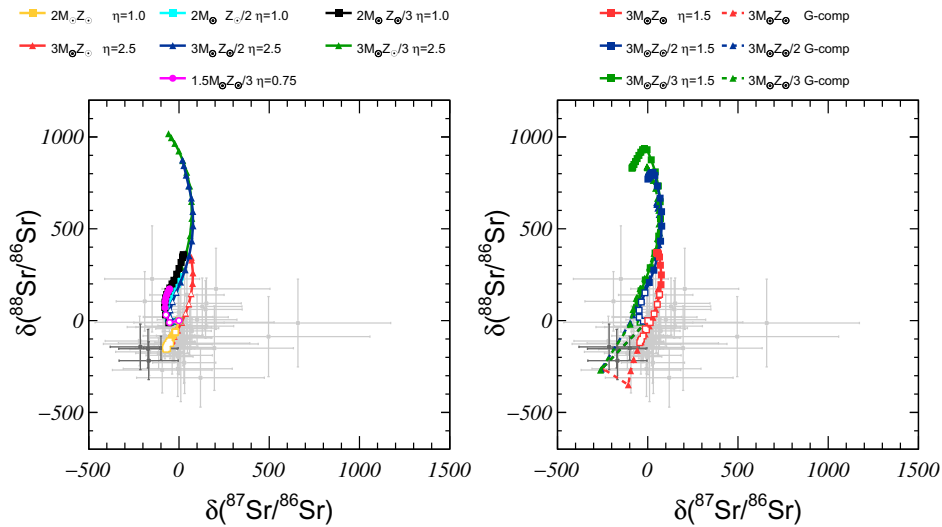

Fig. 4. Comparison between our model predictions and presolar-grain relative isotopic abundances for Sr. The plot reports $\delta$ (per mill) values for the ratios $^{88}$Sr/$^{86}$Sr and $^{87}$Sr/$^{86}$Sr. Data points (full dots) refer to measurements from Liu et al. (2015) and from the WUSTL database (see Hynes and Gyngard, 2009 and subsequent updates). We plot in a darker color the data having error bars smaller than 100% of the measurements themselves. The curves show the evolution in time of the abundances in the stellar envelopes; the dots along the lines represent the various TPs. Open symbols refer to the O-rich phase, solid symbols to the C-rich one. Model calculations are for AGB stars with mass from 1.5 to 3M$_\odot$ and metallicity from 1/3 solar to solar. The choices for the mass loss parameter $\eta$ are indicated. The isotopic shifts relative to the stellar G-component are included to show the maximum effect achievable by nucleosynthesis processes. See discussions in the text. (A color version of this figure is available in the online journal.)

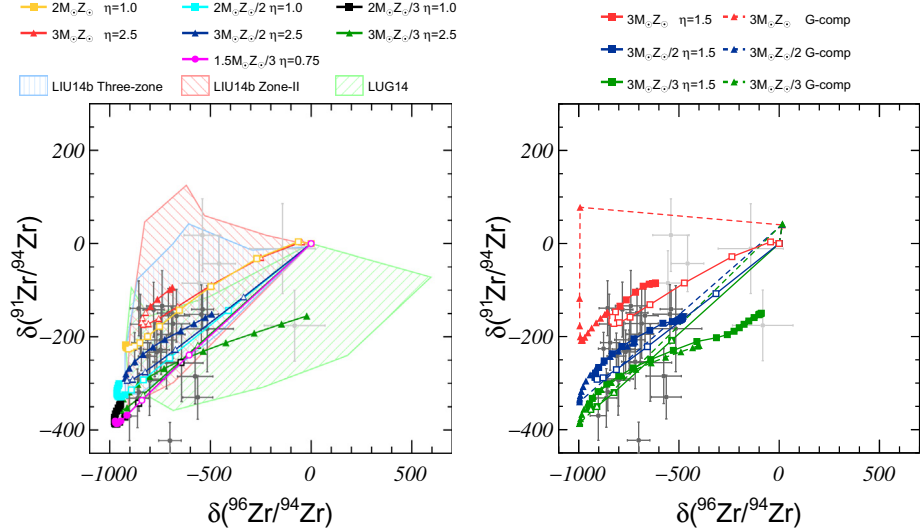

Fig. 5.— An example of a three-isotope plot for Zr. The measurements of $\delta(^{91}Zr/^{94}Zr)$ and $\delta(^{96}Zr/^{94}Zr)$ are from the sources described in the text. Dark dots represent data with error bars smaller than 70% of the measurements themselves. Curves are from the same models of Fig. 4. Shaded areas represent the regions covered by some existing calculations as indicated in the labels. (A color version of this figure is available in the online journal.)

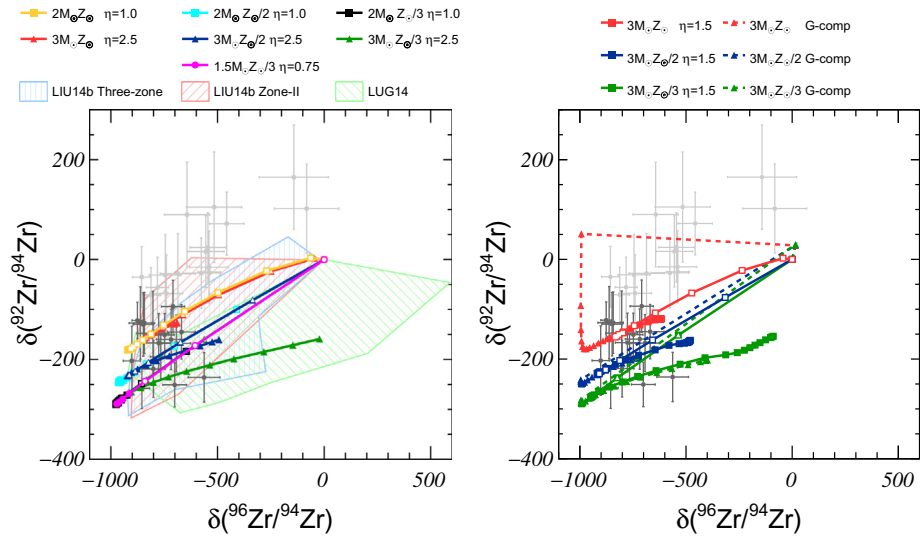

Fig. 6.— Same as Fig. 5, but showing in the ordinate the measurements of $\delta(^{92}Zr/^{94}Zr)$. The same choices have been done for the dark dots and for the shaded areas. (A color version of this figure is available in the online journal.)

We also show zirconium (Figs. 5 and 6) because it is the most typical *ls* element and because its isotopic ratios in SiC grains were extensively discussed in Liu et al. (2014b) and Lugaro et al. (2014). Then we included in Fig. 7 the isotopic ratios of Ba (as already done by Liu et al., 2014a), which is considered as a typical *hs* element. Finally, we added comparisons of the trends displayed by Sr and Zr isotopes with respect to those of Ba (Figures from Figs. 8–10), because Liu et al. (2015) suggested the correlation between *ls* and *hs* to be a crucial constraint on the $^{13}C$ pocket.[2] In order to permit an evaluation of what different choices for mass loss would imply for the isotopic shifts in *s*-process isotope ratios, as anticipated we present in the figures also the model curves pertaining to the "G-component" of a few calculations (those for 3$M_\odot$ stars), indicated by dashed lines. They represent the maximum nucleosynthesis effects possible for each model

We believe the plots present, in general, a rather satisfactory agreement between models and measurements, despite the fact that the error bars (of experimental data) are large and that not all the points are reproduced (see in particular some areas covered by the data, but not by the models, in Figs. 4, 7 and 10). The quality of the agreement between the data and the curves can in any case be evaluated only in comparison with previous attempts at fitting the SiC isotopic abundances, as available in the literature. This comparison is therefore at the base of the discussion presented in Section 5.

---

[2] Similar comparisons for further isotopic and elemental admixtures among those measured are available to interested readers upon request.

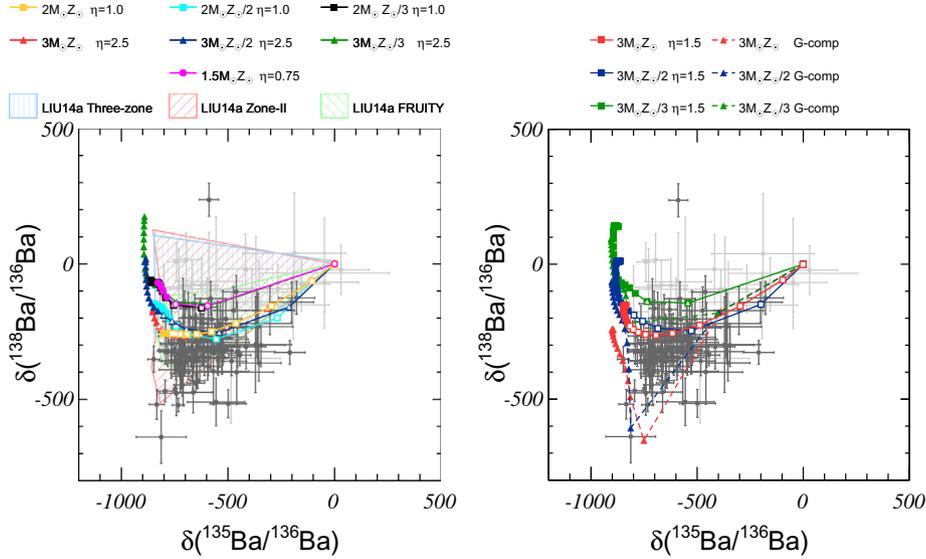

Fig. 7. An example of a three-isotope plot for Ba. The measurements of $\delta(^{138}\mathrm{Ba}/^{136}\mathrm{Ba})$ and $\delta(^{135}\mathrm{Ba}/^{136}\mathrm{Ba})$ are from the WUSTL database and from Liu et al. (2015). Our nucleosynthesis models are the same as in previous figures. The notations used are the same as in Figs. 5 and 6. Shaded areas cover the region of the models in Liu et al. (2014a). (A color version of this figure is available in the online journal.)

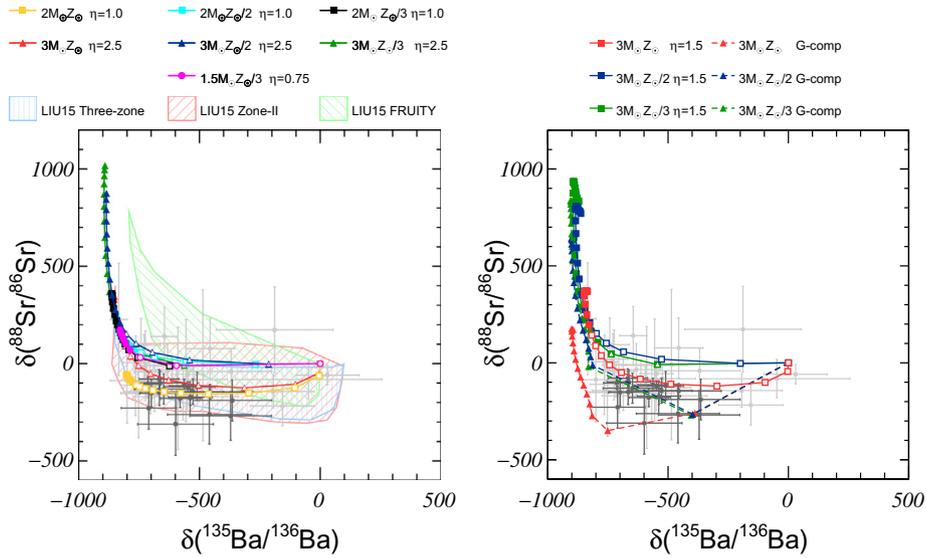

Fig. 8. A comparison between our model predictions and $\delta$ values for $^{88}\mathrm{Sr}/^{86}\mathrm{Sr}$ and $^{135}\mathrm{Ba}/^{136}\mathrm{Ba}$, as presented in Liu et al. (2015) and in the WUSTL database. Dark dots represent data with error bars smaller than 100% of the measurements themselves. Shaded areas display the regions covered by some existing calculations reported by Liu et al. (2015) (see discussion in the text). (A color version of this figure is available in the online journal.)

## 5. DISCUSSION

Besides the Liu et al. (2015) analysis, a few other attempts were made at reproducing the measurements discussed so far making use of a database for neutron capture cross sections comparable to the one we adopt here. They were made using exploratory descriptions for the $^{13}$C pocket, as outlined in the previous sections (see also, e.g., Liu et al., 2014a; Liu et al., 2014b; Lugaro et al., 2014). Our examples presented in Figures from Figs. 4–8 compare well with the best results obtained by the above authors and have the advantage of deriving from a physically-based approach. There are also some significant improvements, revealed by our fits. An example of this concerns strontium. While Liu et al. (2015) could explain the whole range of measurements for $^{88}\mathrm{Sr}/^{86}\mathrm{Sr}$ with their models, the same was not true for $^{87}\mathrm{Sr}/^{86}\mathrm{Sr}$; for it, $\delta$ values in excess of zero could not be reproduced (see Figs. 5 and 7 in that paper). Also in our case, the range for $^{88}$Sr/$^{86}$Sr is accounted for completely, while some of the measurements for $^{87}$Sr/$^{86}$Sr are not fitted: however, the models presented in our Fig. 4 achieve higher δ values than in previous work (about 100, instead of 0). This fact is directly related to the remarkable modifications in the *ls* nuclei obtained in our *s*-process distributions with respect to previous attempts, with a stronger contribution to Sr isotopes coming from the main component of the *s*-process Trippella et al. (2016).

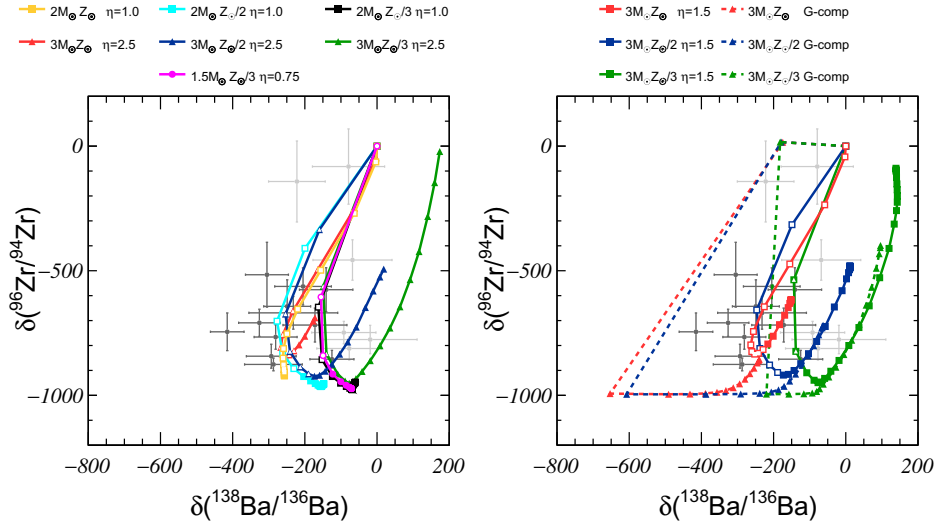

Fig. 9. An example of the comparisons between our model predictions and δ values for some Ba and Zr isotopes, using δ($^{138}$Ba/$^{136}$Ba) as abscissa). Measurements in presolar SiC grains are from the WUSTL database. Dark dots refer to data with uncertainties lower than 70% of the measurements. Model curves and notations are the same as in previous figures. (A color version of this figure is available in the online journal.)

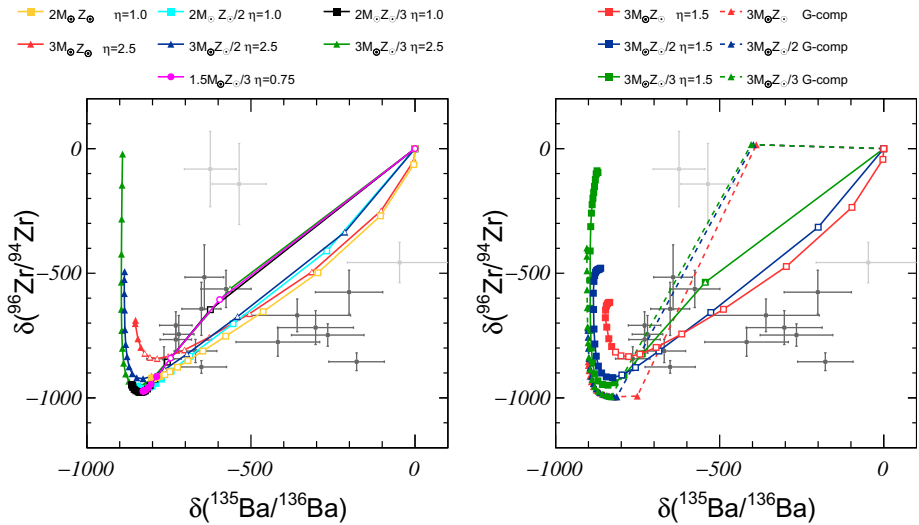

Fig. 10. Another example of the comparisons between our model predictions and δ values for isotopic ratios of Ba and Zr, but using δ($^{135}$Ba/$^{136}$Ba) as abscissa. Model curves, notations are the same as in the previous figure. (A color version of this figure is available in the online journal.)

Another remarkable case concerns zirconium. For it, Lugaro et al. (2014) had difficulties in explaining the highest values of the $^{96}$Zr/$^{94}$Zr and $^{92}$Zr/$^{94}$Zr ratios (see their Figs. 3 and 4, central panels); similarly, Liu et al. (2014b) met difficulties for some low values of $^{91}$Zr/$^{94}$Zr (see the differently shaded areas in our Figs. 5 and 6). These problems are now considerably reduced in our results, albeit not completely solved. The curves referring to the G-component show that improvements on this comparison require to consider cases in which the dilution with the envelope is lower, i.e. where the composition is closer to that of pure *s*-processed material.

As discussed in Liu et al. (2014a), Liu et al. (2014b) and Liu et al. (2015), working inside the scenario of the so-called "Torino group", which is at the base of their work, these authors could not reproduce acceptably the data; they had therefore to make exploratory adjustments of the shape and abundances of the $^{13}$C pocket. In our case, as explained before, the pocket details descend from the original MHD model. whose characteristics were already fixed on different grounds and cannot be further fine tuned (Nucci and Busso, 2014; Trippella et al., 2016; Palmerini et al., 2017). We believe the Figures do show that a rather satisfactory fit is in fact possible, athough perfect agreement is not reached. This comes with the extra property of the model to reproduce also the isotopic abundances of nuclei affected by *p*-captures, typical of evolved LMS (Palmerini et al., 2017), within a unique approach, which seems for the moment to offer a rather general tool for treating deep mixing in evolved red giants.

The results by Liu et al. (2014b) and their discussion on these points clarify that they could improve the agreement with the data by expanding the pocket and flattening the $^{13}$C profile with respect to their original exponential $^{13}$C decay. These are exactly the characteristics of our new $^{13}$C reservoir (see Fig. 1). In our scenario, neutron captures at low neutron density are favored with respect to past attempts due to the extension and shape of the $^{13}$C distribution and this induces non-negligible variations in elemental and isotopeic ratios (Trippella et al., 2016). The present comparison confirms that these variations do permit a better reproduction of the Sr and Zr isotopic ratios than possible before, as illustrated in Figs. 5 and 6.

Other attempts at reproducing the SiC *s*-process data, made previously in the literature, e.g. those by Lugaro et al. (2003), are difficult to consider here, because they were based on a more limited sample of measurements than available now and on older choices for the neutron-capture cross sections.

Fig. 7 shows an example of our predictions for the Ba isotopes. For them we followed the suggestion by Liu et al. (2014a) that the decay of $^{135}$Cs ($t_{1/2} = 2$ Ma) had no time to occur before dust formation in the circumstellar envelope and must therefore be excluded in computing the $^{135}$Ba/$^{136}$Ba ratio of SiC grains. In general, we account for the measurements of Ba isotopes at the same level of accuracy previously possible, considering together the results by Lugaro et al. (2014) and the whole set of different attempots by Liu et al. (2014a). However, some data points still remain outside the reach od AGB modelling. The reasons for this fact need to be explored further, especially when measurements with smaller error bars become available. Perhaps the most important discrepancy refers to Ba isotopes. In particular, the lowermost part of the measured points in Fig. 7, corresponding to the less-negative δ values for $^{135}$Ba/$^{136}$Ba, is not well fitted (neither by us, nor by previous works). The same is true in Fig. 10. We suspect that here we might find interesting indications on the nuclear parameters controlling the nucleosynthesis of *s*-process isotopes. These problems will be therefore reanalyzed, together with Mo data, in the already quoted forthcoming paper, including models for more massive AGB stars and a scrutiny of the nculear uncertainties.

Relevant tests on the extension of the pocket and on the form of the $^{13}$C distribution can then be obtained by the relative trends of *ls* and *hs* nuclei, as done by Liu et al. (2015) for Sr and Ba.

Due to the above reasons, in Fig. 8 we present, for the correlation between Sr and Ba, a first comparison of our results with those by Liu et al. (2015), including model predictions cited by them (taken either from their own calculations or from the FRUITY on-line database). Again, results from those previous works are shown in the figure as shaded areas. As the plot illustrates, our model curves can account for almost the whole area covered by the data points, something that was possible in the mentioned paper only by varying in exploratory ways both the size of the pocket and the abundances of $^{13}$C and $^{14}$N in it. The good match between models and measurements, which we find with our rather extended and flat $^{13}$C distribution, is in any case another indication that the general characteristics of the $^{13}$C pocket outlined by the above authors were correct.

An equally important test should be provided by plots of Zr isotopes versus Ba ones and we present two examples of them in Figs. 9 and 10. Unfortunately, we cannot make direct comparisons with previous models. Indeed, the comparisons between the ratios we show were not addressed by Liu et al. (2015). We leave therefore our model curves as predictions of the range we expect and of the quality of our fits, for consideration in future works.

Thanks to the analysis performed by Liu et al. (2015), we can infer that the satisfactory result we obtain for Sr and Ba isotopes is due to the fact that our $^{13}$C reservoir has the general characteristics indicated by them as necessary (rather flat $^{13}$C distributions, very low $^{14}$N abundances and a large extension).

We also hope to have found a specific physical mechanism suitable to account for the constraints from presolar SiC grains, because our fits are in general of a quality at least comparable to the best ones discussed in Liu et al. (2014a), Liu et al. (2014b), Liu et al. (2015) and Lugaro et al. (2014).

## 6. CONCLUSIONS

In this paper we presented predictions for the isotopic admixture of *s*-elements, as resulting by the assumption that the main neutron source $^{13}$C be formed by MHD processes. Our predictions were compared with the existing record of isotopic and elemental measurements for *n*-capture elements in presolar SiC grains recovered from pristine meteorites. Our fits were obtained starting from a general mathematical solution of the MHD equations. This was implemented into a physically-meaning approach in other contexts (the explanation of $^{26}$Al and of the isotopic abundances of oxygen and carbon in presolar oxyde grains and the reproduction of solar and post-AGB abundances). This was done by fixing the two parameters $v_{r,0}$ and $B_{\varphi,0}$ to values compatible with observations. Here we simply inherit those findings, without adding further adjustable terms: the $^{13}$C reservoir (Fig. 1) is obtained as a necessary consequence and turns out to have characteristics very similar to those suggested by Liu et al. (2015).

This also confirms the correctness of the indications by those authors: not only did we verify that these characteristics permit to reproduce the data, but also that they can be actually obtained in a physical model. On the other side, our work shows that the *s*-process scenario depicted by Trippella et al. (2016) is flexible enough to account for observational and experimental constraints coming from heterogeneous sources.

Quite obviously, our discussion does not imply that the above model is "unique" in explaining the SiC grain composition. Various mechanisms based on hydrodynamic or magneto-hydrodynamic mixing processes having a similar speed (10–100 m/s) might be equally good, if they can yield a $^{13}$C profile flat and extended enough to reproduce the constraints suggested by Liu et al. (2015). With this analysis we therefore aim also at stimulating the authors that are exploring other physical mechanisms (e.g. gravity waves, or revisions of the physics at the convective envelope border) to repeat our analysis. We trust this would provide current research with new useful indications on the general problem of extra-mixing in evolved stars.

ACKNOWLEDGEMENTS

This work was partially supported by the Department of Physics and Geology of the University of Perugia, under the grant named "From Rocks to Stars", and by INFN, section of Perugia, through funds provided by the Scientific Commission n. 3 (Nuclear Physics and Astrophysics). O.T. is grateful to both these organizations for post-doc contracts. S.P. and A.Z. acknowledge the support of Fondazione Cassa di Risparmio di Perugia. M.P. acknowledges the financial support of the European Research Council for the Consolidator Grant ERC-2013–383 CoG Proposal No. 612776 CHRONOS

*Associate editor:* Anders Meibom